\pdfoutput=1
\documentclass[sigconf]{acmart}

\usepackage{booktabs} % For formal tables


\copyrightyear{2018} 
\acmYear{2018} 
\setcopyright{acmcopyright}
\acmConference[CIKM '18]{The 27th ACM International Conference on Information and Knowledge Management}{October 22--26, 2018}{Torino, Italy}
\acmBooktitle{The 27th ACM International Conference on Information and Knowledge Management (CIKM '18), October 22--26, 2018, Torino, Italy}
\acmPrice{15.00}
\acmDOI{10.1145/3269206.3272020}
\acmISBN{978-1-4503-6014-2/18/10}

\usepackage[T1]{fontenc}
\usepackage{epstopdf}
\usepackage{multirow}
\usepackage{balance}
\usepackage{amsfonts}
\usepackage{amsmath}
\usepackage{amssymb}
\usepackage{graphicx}
\usepackage{subfigure}
\usepackage[noend,ruled]{algorithm2e}
\usepackage{microtype}
\usepackage{url}
\usepackage{enumitem}

\usepackage{tabularx, booktabs}

\newtheorem{thm:def}{Definition}
\newtheorem{thm:eg}{Example}
\newtheorem{thm:lem}{Lemma}
\newtheorem{thm:obs}{Observation}
\newtheorem{thm:conj}{Observation}
\newtheorem{thm:req}{Requirement}

\newcommand{\nop}[1]{}

\newcommand{\ie}{{\sl i.e.}}
\newcommand{\eg}{{\sl e.g.}}

\newcommand{\our}{\mbox{\sf Maester}\xspace}
\newcommand{\method}[1]{\mbox{\sf #1}\xspace}

\newcommand{\X}{\mathbf{X}}
\newcommand{\W}{\mathbf{W}}
\newcommand{\V}{\mathbf{V}}
\newcommand{\w}{\mathbf{w}}
\newcommand{\m}{\mathbf{m}}
\newcommand{\x}{\mathbf{x}}
\newcommand{\fg}{\mathbf{f}}
\newcommand{\ig}{\mathbf{i}}
\newcommand{\og}{\mathbf{o}}
\newcommand{\mc}{\mathbf{c}}
\newcommand{\mb}{\mathbf{b}}
\newcommand{\h}{\mathbf{h}}
\newcommand{\att}{\mathbf{a}}

\newcommand{\smallsection}[1]{\vspace{1mm}\noindent\textbf{#1.}}	% define own new subsection type: noindent, bold

\DeclareMathAlphabet{\mathbbold}{U}{bbold}{m}{n}

\begin{document}

% \title{Breaking Filter Bubbles for Opinion Questions}
\title{Investigating Rumor News Using Agreement-Aware Search}

\author{Jingbo Shang$^1$, Jiaming Shen$^1$, Tianhang Sun$^1$, Xingbang Liu$^1$,\\ Anja Gruenheid$^2$, Flip Korn$^3$, \'Ad\'am D. Lelkes$^3$, Cong Yu$^3$, Jiawei Han$^1$}
\affiliation{%
  \institution{$^1$Department of Computer Science, University of Illinois Urbana-Champaign, IL, USA}
}
\affiliation{%
  \institution{$^2$Google Inc., Madison, WI, USA $\quad$ $^3$Google Research, New York, NY, USA}
}
\affiliation{%
  \institution{$^1$\{shang7, js2, ts7, xl14, hanj\}@illinois.edu $\quad$ $^{2,3}$\{anjag, flip, lelkes, congyu\}@google.com}
}

\renewcommand{\shortauthors}{J. Shang et al.}

\begin{abstract}
    % !TEX encoding = UTF-8
%!TEX root = 00-main.tex

Recent years have witnessed a widespread increase of rumor news generated by humans and machines.
% in order to attract readership, influence opinions, and increase click-through revenue. 
Therefore, tools for investigating rumor news have become an urgent necessity.
One useful function of such tools is to see ways a specific topic or event is represented by presenting different points of view from multiple sources.
In this paper, we propose \our, a novel agreement-aware search framework for investigating rumor news.
Given an \emph{investigative question}, \our will retrieve \emph{related} articles to that question, assign and display top articles from \emph{agree}, \emph{disagree}, and \emph{discuss} categories to users.
Splitting the results into these three categories provides the user a holistic view towards the investigative question.
We build \our based on the following two key observations:
(1) relatedness can commonly be determined by keywords and entities occurring in both questions and articles,
and (2) the level of agreement between the investigative question and the related news article can often be decided by a few key sentences.
Accordingly, we use gradient boosting tree models with keyword/entity matching features for relatedness detection, and leverage recurrent neural network to infer the level of agreement.
Our experiments on the Fake News Challenge (FNC) dataset demonstrate up to an order of magnitude improvement of \our over the original FNC winning solution, for agreement-aware search. 

\end{abstract}


\settopmatter{printacmref=false, printfolios=false}

\maketitle

{\fontsize{8pt}{8pt} \selectfont
\textbf{ACM Reference Format:}\\
Jingbo Shang, Jiaming Shen, Tianhang Sun, Xingbang Liu, Anja Gruenheid, Flip Korn, Adam D. Lelkes, Cong Yu, Jiawei Han. 2018. Investigating Rumor News Using Agreement-Aware Search. In The 27th ACM Int'l Conference on Information and Knowledge Management (CIKM'18), Oct. 22--26, 2018, Torino, Italy. ACM, New York, NY, USA, 9 pages. \url{https://doi.org/10.1145/3269206.3272020}
}

%!TEX root = 00-main.tex
%!TEX encoding = UTF-8

\vspace{-0.2cm}
\section{Introduction}\label{sec:intro}

Increasing amounts of rumor news have been generated and widely spread in recent years, in order to attract readership, influence opinion, and increase click-through revenue.
This is a serious problem for the news industry as unreliable news increases mistrust of the media and may have wide-reaching implications such as impact on elections~\cite{tavernise2016fake,connolly2016fake}.
According to a research poll, 64\% of US adults say that rumor news has caused a ``great deal of confusion'' about the factual content of reported current events~\cite{barthel2016many}.
Therefore, tools for investigating rumor news have become an urgent necessity.

\begin{figure}[t]
  \centering
  \includegraphics[width=0.9\columnwidth]{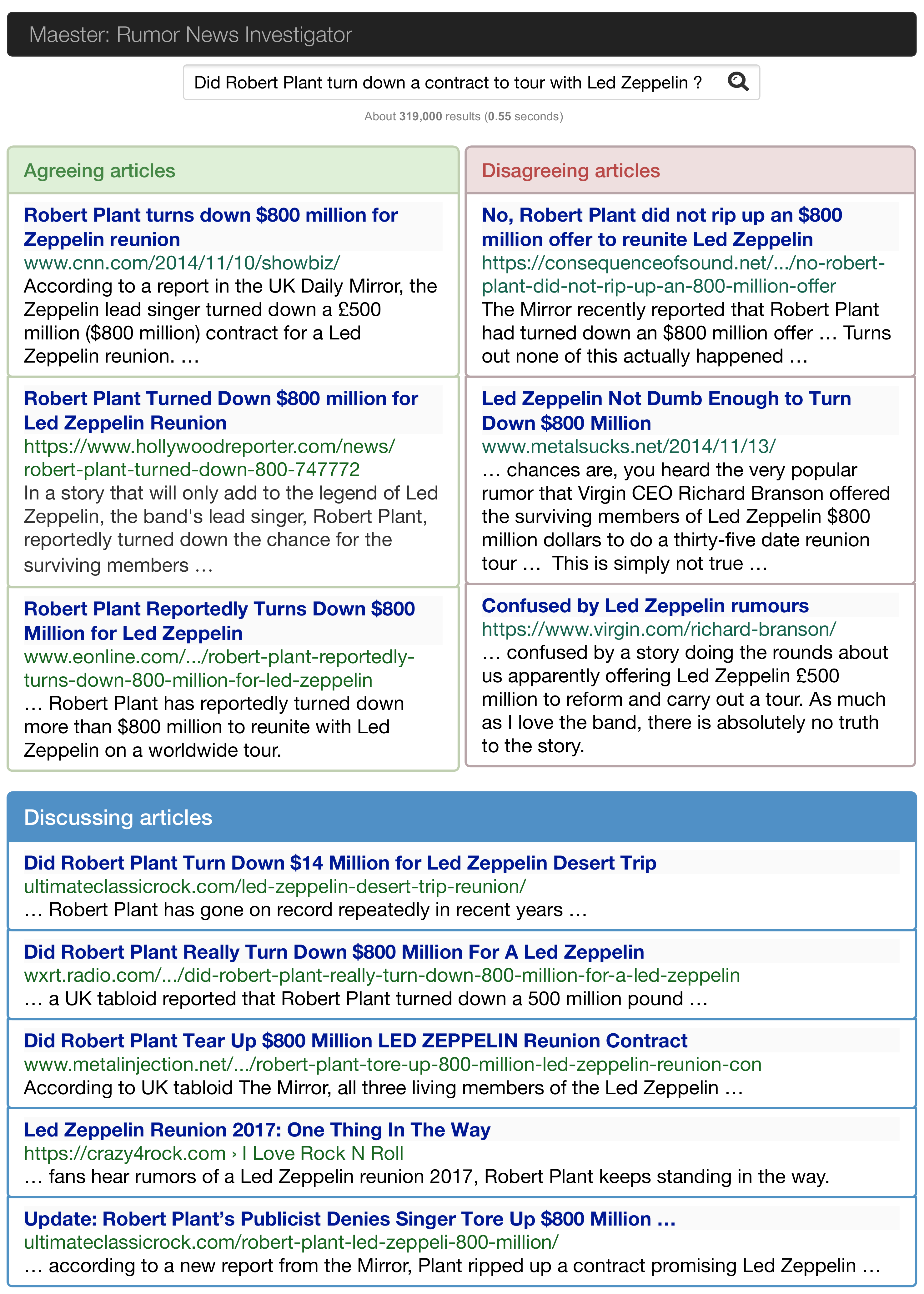}
  \vspace{-0.3cm}
  \caption{The interface of our proposed agreement-aware search framework, \our. Instead of a traditional ranked list of \emph{related} articles, we propose to present 3 \emph{agree} articles, 3 \emph{disagree} articles, and 5 \emph{discuss} articles respectively for a given investigative question.}\label{fig:interface}
  \vspace{-0.4cm}
\end{figure}

One useful function for such tools is to see ways a specific topic or event is represented by presenting different points of view from multiple sources.
Often, these topics can be phrased as \emph{investigative questions} such as our running example, ``Did Robert Plant turn down a contract to tour with Led Zeppelin?''
For this question, some news articles reported Robert Plant turned down the contract while others disputed that it was not true; yet others merely summarized an existing article without stating its own position.
In this sense, this question could be considered \emph{controversial}.
Such function is beneficial to not only users but also specialists like a journalist working on a fact-checking article or a historian cataloging beliefs and trends.

In this paper, we study how to automatically identify the stances of news articles and rank them based on their levels of agreement with a given question. 
Specifically, we propose \our, a novel agreement-aware search framework. 
Given an investigative question, \our will first retrieve \emph{related} articles that address the target question. 
Each of these articles is then automatically assigned a stance label of either \emph{agree}, \emph{disagree}, or \emph{discuss},
where \emph{discuss} pertains to articles that merely discuss or summarize other articles reporting on the reference question without making a statement of their own with regard to the question.
Splitting the results into these three categories allows the user to
(a) see quickly whether a topic is controversial (e.g., some category does not have any assigned articles),
(b) get an overview of the different points of view, and
(c) form a more informed understanding about the sources taking a position and evidence presented in the articles.

Our methodology is based on the following two observations from real-world rumor news articles:
(1) relatedness of an article can often be determined by its shared keywords/entities with the investigative question; and
(2) agreement level of an article can often be inferred from a few key sentences in it.
For example, as shown in Figure~\ref{fig:interface}, all retrieved articles are related through the keywords ``Robert Plant'' and ``Led Zeppelin'', and we can determine their stances based on the sentences shown in the search result snippets. 
Accordingly, we design \our as a two-step framework, which first filters unrelated articles and then predicts agreement status of remaining related articles. 
We learn a gradient boosting tree model with four types of features, including the key entity features, to classify whether an article is related to question or not. 
Then, we select top-3 sentences in each \emph{related} article that are closely correlated to the investigative question.
These sentences, together with the reference question, are then fed into a recurrent neural network (RNN) which outputs the level of agreement for each news article. 
Finally, \our ranks these news articles and displays top-ranked ones within each agreement category to users. 

We evaluate \our using the dataset from the Fake News Challenge\footnote{http://www.fakenewschallenge.org/} (FNC).
Extensive experiments 
verify our two observations empirically
and 
demonstrate the significant improvements of \our over the original challenge winner's solution (i.e.,~an ensemble model of gradient boosting trees and a convolutional neural network).
In summary, our contributions are as follows.
\begin{itemize}[leftmargin=*,noitemsep]
    \item {\bf Agreement-Aware Search Framework.} We propose and build a novel agreement-aware search framework, \our, to bring a holistic view to the user towards the investigative question.
    \item {\bf Agreement Detection.} We propose a novel model based on RNN with attention mechanism for classifying and ranking related articles by stance.
    \item {\bf Extensive Evaluation.}
    We conduct a thorough experimental evaluation to demonstrate the effectiveness of \our by comparing it with the FNC first-place method.
    For controversial questions, \our achieves a significant improvement for overall agreement-aware ranking ($\sim$2x), with a 7-fold improvement in the especially difficult case of disagreement;
    over both controversial and non-controversial questions, the improvement is $20\%$.
    In addition, it improves over the first-place method in terms of the FNC weighted accuracy metric by $2.88\%$.
\end{itemize}

% The remainder of this paper is organized as follows.
% The related work is discussed first in Section~\ref{sec:rel}.
% Then, the problem formulation and framework design are introduced in Section~\ref{sec:pre}.
% Section~\ref{sec:method} covers the technical details of our proposed framework and Section~\ref{sec:exp} contains extensive experiments testing that framework on real-world data.
% We conclude the study and discuss some future directions in Section~\ref{sec:con}.

\begin{figure*}[t]
  \centering
  \includegraphics[width=0.8\textwidth]{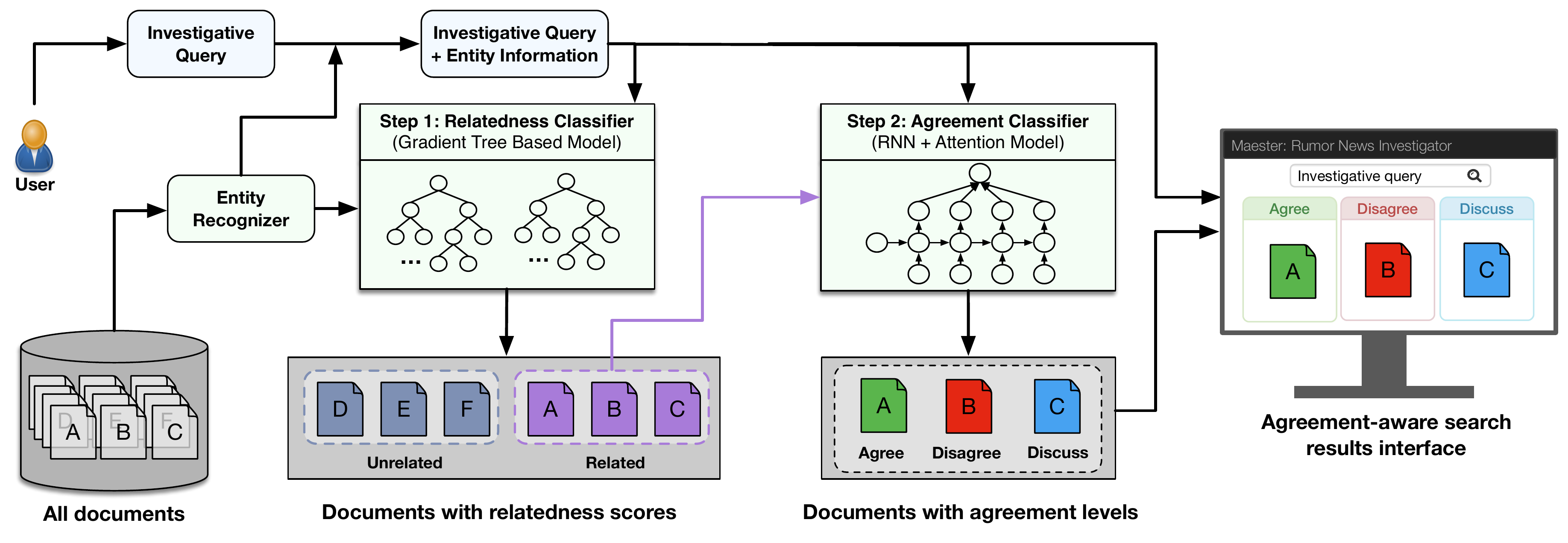}
  \vspace{-0.5cm}
  \caption{Overview of \our framework.}\label{fig:overview}
  \vspace{-0.5cm}
\end{figure*}

% !TEX encoding = UTF-8
%!TEX root = 00-main.tex

\vspace{-0.3cm}
\section{Related Work}\label{sec:rel}

% In this section, we will review available literature related to agreement detection of news articles, question answering and other lines of work relevant to our problem.
In this section, we review literature related to agreement detection of news articles, question answering, and other lines of work relevant to our studied problem.

\smallsection{Stance Detection}
The natural language processing community has explored stance detection for years and have formulated it in various ways.
\emph{SemEval 2016 Task 6} defines it as determining from text whether the author is in favor of, against, or neutral towards a given target~\cite{Mohammad2016SemEval2016T6}.
In this shared task, the text is a tweet and the target is a single entity without any descriptive text.
Following the same line of work, researchers have explored how to decide whether a tweet or an article favors one specific entity over others~\cite{somasundaran2009recognizing}.
However, finding agreement with respect to an investigative question is more challenging than simply determining the stance for specific entities.
This is because any subtle changes in the wording may lead to a completely different interpretation of the question.

Mohammad et al.\ first released a dataset for tweet stance~\cite{mohammad2016dataset}, and later studied sentiment and stance for tweets~\cite{MohammadSK17}.
Other approaches to stance detection in social media include semi-supervised topic models to classify stance~\cite{xu2017semi} and latent feature extraction~\cite{xu2017efficient}.
Furthermore, stance detection has been explored in Chinese microblogs~\cite{Xu2016OverviewON} and online discussion forums~\cite{skeppstedt2017automatic}.
All of these tasks require exactly one targeted entity, however, investigative questions may contain more than one entity.
Thus, these methods cannot be directly adopted for our use case.

\smallsection{Agreement Detection in FNC-1}
In the summer of 2017, the \emph{Fake News Challenge} (FNC) ran its first contest on agreement detection.
The task of this contest was to determine agreement given pairs of headlines and news articles.
The challenge provides a partially labeled dataset, denoted in the following as \emph{FNC-1}, which is based on the \emph{Emergent} dataset~\cite{Ferreira2016EmergentAN}, and contains rumor news.
The winner of the FNC-1~\cite{fnc-1-winner} developed an ensemble model of a tree-based model and a CNN-based model.
Similar to the solution to rumor news detection proposed in this work, the tree-based model utilizes a set of handcrafted features, however, it neglects important entity features.
The CNN-based model on the other hand can extract features automatically but its performance is not as good as that of the tree-based model.
We use the FNC-1 dataset for our evaluation and compare \our with the winner's solution in Section~\ref{sec:exp} thoroughly.
Note that all challenge winners~\cite{zarrella2016mitre,wei2016pkudblab,fnc-1-winner} in SemEval and FNC take advantage of both handcrafted and neural network based features.
\our also follows the same paradigm.

\smallsection{Textual Entailment}
Another related line of work is textual entailment, which studies whether a text entails, contradicts, or not related to a certain hypothesis~\cite{androutsopoulos2010survey,wang2015learning,rocktaschel2015reasoning}.
However, entailment emphasizes the logical relation of text and hypothesis where the text is commonly only one sentence and thus is much shorter than a news article.

\smallsection{Question Answering}
Question answering (QA) is the task of finding an article, a passage, or a sentence to answer a given question~\cite{voorhees1999trec}.
Most, if not all, of these questions have a specific and clear answer.
However, this work focuses on controversial questions for which traditional question answering systems do not work well.
For example, given a simple fact-seeking question like ``Was George Washington a U.S. president?'' one should only find \emph{agree} articles.
In contrast, controversial questions lack consensus and often become a hotbed for spreading rumor news.\footnote{
We recognize the sensitivity and importance of not propagating conspiracy theories (e.g.,~``Did 9/11 really happen?'') and, for now, propose to deal with this challenge by limiting candidate results to trusted sources.
}% end footnote
As a result, traditional QA systems struggle to address this modified problem.

\smallsection{Search Result Diversification}
Search result diversification~\cite{drosou2010search} has been originally proposed to deal with query ambiguity, and has been applied to improve personalized search~\cite{radlinski2006improving} afterwards.
In the same context, query reformulation~\cite{santos2010exploiting} has been explored to retrieve more relevant articles per target, and thus diversifying the search results. In \cite{dang2012diversity}, the authors furthermore propose to consider the proportionality of articles instead of emphasizing diversity.
However, depending on the diversity measure, articles within the same agreement group can also be diverse.
Therefore, directly applying search diversification methods cannot guarantee the presence of all agreement groups.
As showing multiple ranked lists for different agreement groups essentially enforces the results to be diversified, we may also apply similar techniques to optimize the overall quality of the ranked lists per agreement group.

% !TEX encoding = UTF-8
%!TEX root = 00-main.tex

\section{Preliminaries}\label{sec:pre}

    In this section, we will first formulate the problem and then discuss our framework design and alternative models.

    \subsection{Problem Formulation}
        Given a question $q$, we assume that a collection of candidate articles $\mathcal{D}(q)$ is provided.
        There are many ways to obtain such a collection (\eg, taking the top-100 articles from a collection based on BM25 scores), which is not the focus of this paper.
        % however, article collection is not the focus of this paper.

        \begin{thm:def}[Agreement Classes]
        Given an investigative question $q$ and an article $d \in \mathcal{D}(q)$, we define four possible classes to describe how $d$ relates to $q$:
        \vspace{-0.1cm}
        \begin{enumerate}[leftmargin=*,noitemsep]
            \item \emph{\textbf{Agree:}} The article agrees with $q$
            \item \emph{\textbf{Disagree:}} The article disagrees with $q$
            \item \emph{\textbf{Discuss:}} The article discusses the same question, but does not take a position w.r.t.\ $q$
            \item \emph{\textbf{Unrelated:}} The article addresses a question other than $q$.
        \end{enumerate}
        \end{thm:def}
        
        Previously, we have noted that the key to rumor detection is to find those questions that lead to controversial discussion of a topic, i.e.,~on which people have more than one opinion. 
        More formally, we use the following definition for controversial questions.

        \begin{thm:def}[Controversial Question]
        When an investigative question has at least one agreeing and one disagreeing news article in $\mathcal{D}(q)$, we refer to it as a controversial question.
        \end{thm:def}

        For understanding controversial questions and agreement classes, consider the following example taken from the FNC that shows text snippets referencing the running example question ``Did Robert Plant turn down a contract to tour with Led Zeppelin?''.
        Here, the controversial question leads to different news articles that can be categorized according to statements made in those articles.
     
        \begin{thm:eg}\label{eg:stance}
        \vspace*{-1ex}
        The running example showing relatedness classification and agreement detection for question ``Did Robert Plant turn down a contract to tour with Led Zeppelin?''
        \emph{
            \begin{table}[!h]
            \vspace{-0.3cm}
                \small
                \begin{tabular}{cp{6cm}}
                \toprule
                Question & Did Robert Plant turn down a contract to tour with Led Zeppelin?\\
                \midrule
                \midrule
                \emph{Agree} & $\ldots$ Led Zeppelin's Robert Plant turned down $\pounds$500 MILLION to reform supergroup. $\ldots$\\
                \midrule
                \emph{Disagree} & $\ldots$ No, Robert Plant did not rip up an \$800 million deal to get Led Zeppelin back together. $\ldots$\\
                \midrule
                \emph{Discuss} & $\ldots$ Robert Plant reportedly tore up an \$800 million Led Zeppelin reunion deal. $\ldots$\\
                \midrule
                \emph{Unrelated} & $\ldots$ Richard Branson's Virgin Galactic is set to launch SpaceShipTwo today. $\ldots$ \\
                \bottomrule
                \end{tabular}
            \vspace{-0.4cm}
            \end{table}
        }
        \end{thm:eg}
        % \vspace*{-.9ex}
        
% \begin{table*}[t]
%     \center
%     \caption{FNC-1 Dataset Statistics.}\label{tbl:dataset}
%     \begin{tabular}{|c|c|c|c|c|c|c|c|c|}
%     \hline
%              & \multicolumn{2}{|c|}{Investigative Questions} & News Articles & \multicolumn{5}{|c|}{Labeled Pairs} \\
%     \hline
%              & All & Controversial  & Total & Total & Unrelated & Discuss & Agree & Disagree\\
%     \hline
%     Training & 1,648     & 260  & 1,683    & 49,972         & 73.13\%   &  17.83\%  & 7.36\%  & 1.68\%\\
%     \hline
%     Testing  & 894       & 211  & 904      & 25,413        & 72.20\%   & 17.57\%   & 7.49\%  & 2.74\%\\
%     \hline
%     \end{tabular}
% \end{table*}

\begin{table*}[t]
    \center
    \caption{FNC-1 Dataset Statistics.}\label{tbl:dataset}
    \vspace{-0.3cm}
    \begin{tabular}{ccccccccc}
    \toprule
             & \multicolumn{2}{c}{Investigative Questions} & News Articles & \multicolumn{5}{c}{Labeled Pairs} \\
    \cmidrule{2-9}
             & All & Controversial  & Total & Total & Unrelated & Discuss & Agree & Disagree\\
    \midrule
    Training & 1,648     & 260  & 1,683    & 49,972         & 73.13\%   &  17.83\%  & 7.36\%  & 1.68\%\\
    \midrule
    Testing  & 894       & 211  & 904      & 25,413        & 72.20\%   & 17.57\%   & 7.49\%  & 2.74\%\\
    \bottomrule
    \end{tabular}
    \vspace{-0.3cm}
\end{table*}
        
        \smallsection{Formal Problem Definition}
        Our goal is to declare whether a candidate news article is related to an investigative question and, if so, how it is positioned w.r.t. that question.
        More formally, we say that $\forall q \in \mathcal{Q}$ and $d \in \mathcal{D}(q)$, there is a label $y \in \{$unrelated, discuss, agree, disagree$\}$ that describes the relationship between $q$ and $d$.
        Note that it is possible that, for a given reference question, any agreement class may contain multiple news articles.
        Therefore, we desire the output of the agreement identification step to be ranked lists per class as shown in Figure~\ref{fig:interface},
        with $k_{agree}$ \emph{agree} articles, $k_{disagree}$ \emph{disagree} articles, and $k_{discuss}$ \emph{discuss} articles, for example, $(k_{agree}, k_{disagree}, k_{discuss}) = (3,3,5)$ as shown in the running example.
        To measure whether an article is related or unrelated, we determine a confidence score $rel(q,d) \in [0,1]$ where a $0$ signifies that $q$ and $d$ are unrelated and $1$ that $d$ is highly related to $q$.
        For related articles, their levels of agreement can be predicted by a classifier that maps an agreement score $\beta(q,d)$ to range from -1 to +1.
        Here $-1$ indicates maximum disagreement and $+1$ indicates maximum agreement.
        Our models then estimate $P(y | q, d)$ for ranking, where (1)~$P(y | q,d) = \beta(q,d)$ holds for agreeing articles, (2)~$P(y | q,d) = -\beta(q,d)$ holds for disagreeing articles, and (3)~$P(y | q,d) = rel(q,d)$ holds for discussing articles.
        For each $d \in \mathcal{D}(q)$, we define its agreement $\hat{y}$ as $\arg\max_{y} P(y | q, d)$.
        Thus, $\hat{y}$ and the corresponding $P(\hat{y} | q, d)$ determine the membership and ranking of an article $d$ w.r.t.\ $q$ in these three lists.

        \smallsection{Model Training \& Evaluation}
        To train our models, we use a training set containing labels for question-article pairs as labeled above.
        After the models have been trained, they are evaluated on a separate set of questions and their candidate articles, as same as the training and verification methodology applied in the FNC.
        This process holds for both, classification and ranking, tasks.

    \subsection{Framework Overview}
        Figure~\ref{fig:overview} presents an overview of our proposed \our framework.
        We structure our approach in two steps analogous to the two problems discussed above, i.e.,~(1) whether an article is \emph{related} to a given question; and (2) predicting a related article's agreement w.r.t.\ the question.
        Intuitively, the actual modeling challenges for these two problems are substantially different.
        We observe that content words and entity mentions in both the given question and the article may play important roles in predicting their relatedness.
        That is, if the article discusses the same or similar set of entities, they should be related.
        \begin{thm:conj}[Relatedness: Keywords and Entities.]\label{conj:relatedness}
            Overlapping keywords and entities between the given question $q$ and a news article $d$ are crucial for determining their relatedness.
        \end{thm:conj}

        In contrast, overlapping entities are weak signals for finding the level of agreement w.r.t.\ a question.
        Specifically, either an \emph{agree} article or a \emph{disagree} article might contain a large number of overlapping keywords and entities.
        Instead, for the task of agreement detection, non-entity words such as adjective, adverbs, and negation words are more important.
        Furthermore, inspired by many examples such as Figure~\ref{fig:interface} and the running example in Section~\ref{sec:pre}, we observe that only a few sentences, referred as {\emph key sentences}, in an article will often reflect the stance w.r.t.\ a given question, especially for news articles.
        For example, from the sentence ``No, Robert Plant did not rip up an \$800 million deal to get Led Zeppelin back together.'' one can easily derive that this article disagrees with the question ``Did Robert Plant turn down a contract to tour with Led Zeppelin?''.
        Thus, we propose our second observation as follows.
        \begin{thm:conj}[Agreement: Key Sentences.]
        \label{conj:stance}
            An article's agreement w.r.t.\ a given question $q$ is largely decided based on a few key sentences. This is due to the ``inverted pyramid'' structure that journalists often follow when writing a news story~\cite{po2003news}.
        \end{thm:conj}

        \begin{figure}[t]
            \centering
            \includegraphics[width = 0.9\columnwidth]{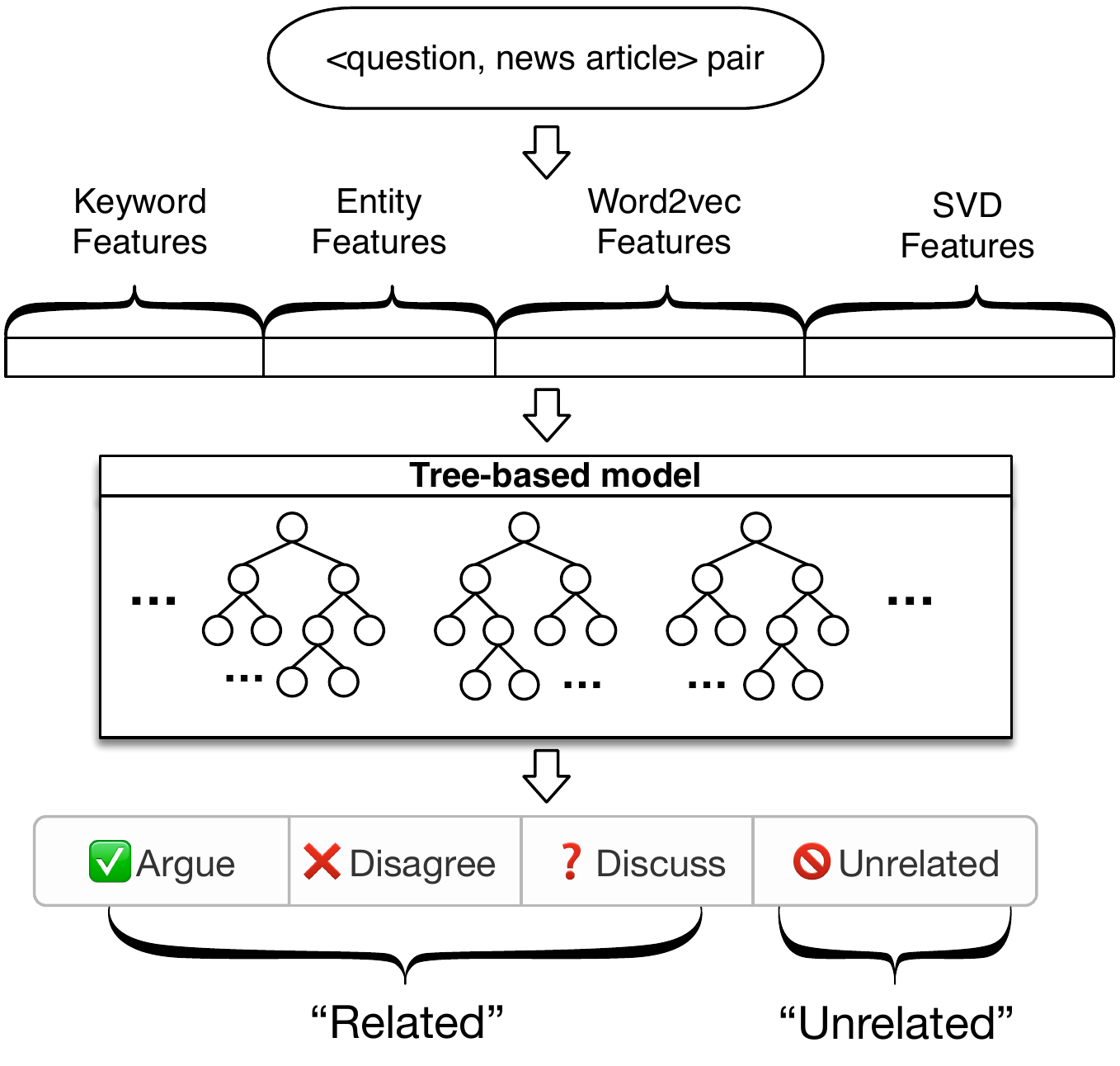}
            \vspace{-0.4cm}
            \caption{Tree-based Classification.}
            \vspace{-0.5cm}
            \label{fig:stage-1}
        \end{figure}

        Finally, we observe that in practice, the distribution of agreement labels is often skewed.
        As shown in Table~\ref{tbl:dataset} for the FNC-1 dataset, the majority of labels are \emph{unrelated} whereas \emph{disagree} has the least number of annotations.
        Avoiding overemphasis of \emph{unrelated} news articles further motivates the following two-step framework.
        \begin{enumerate}[leftmargin=*,noitemsep]
            \item \textbf{Relatedness Classification}.
                First, we merge the four stances into two categories, i.e., \emph{related} and \emph{unrelated}, and focus on the binary classification.
                Based on Observation~\ref{conj:relatedness}, for a given question and an article, we design keyword, entity, word2vec, and SVD features based on the keywords and entity mentions.
                Taking these features as input, as shown in Figure~\ref{fig:stage-1}, our tree-based model leads to a test accuracy close to $98\%$ in our experiments, which verifies this observation empirically.
            \item \textbf{Agreement Detection}.
                Second, for all \emph{related} articles, we build a 3-class classification model to estimate the agreement class.
                Inspired by Observation~\ref{conj:stance}, for a given question and an article, we project the question and every sentence of the article into the embedding space and then choose the most similar sentences as key sentences.
                Afterwards, we inject these sentences into an efficient RNN model with attention mechanism.
                Note that if we instead train a tree-based model using the same keyword/entity-based handcrafted features designed for relatedness classification, the performance drops significantly which is consistent with our observation.
        \end{enumerate}

%!TEX root = 00-main.tex
%!TEX encoding = UTF-8

    \begin{figure*}[t]
        \centering
        \includegraphics[width = 0.9\textwidth]{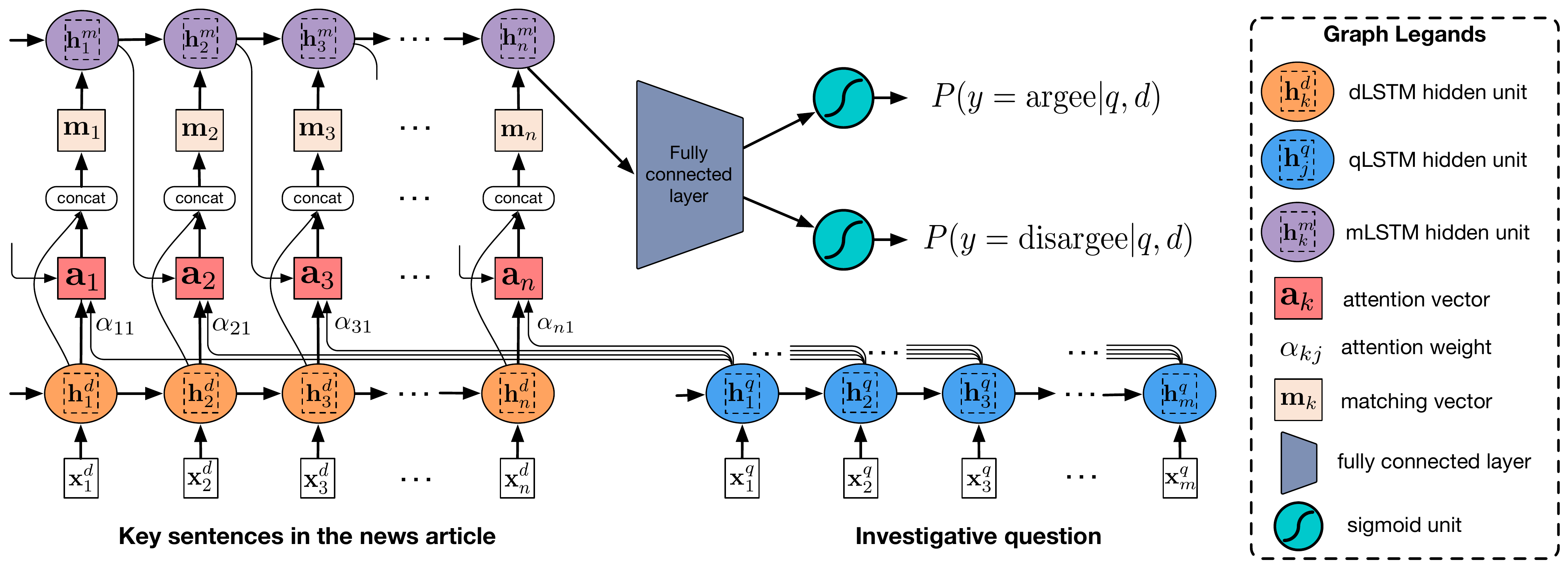}
        \vspace{-0.3cm}
        \caption{The architecture of our proposed RNN+attention Model.}
        \vspace{-0.2cm}
        \label{fig:rnn}
    \end{figure*}

\section{Methodology}\label{sec:method}

    This section first introduces our feature design for the tree-based model which is used to compute relevance scores.
    Then, we present our RNN model with attention mechanism.

    \subsection{Relatedness Classification}
        In this section, we briefly introduce the features used in the relatedness classification.
        As shown in Figure~\ref{fig:stage-1}, we design the following features for each question-article pair and categorize them into four different types:
        (1) keyword features, (2) entity features, (3) word2vec features, and (4) SVD features.

        \smallsection{Keyword Features}
        We compute the non-stopword keyword overlap between the question $q$ and the news article $d$, i.e., $|q \cap d| = \sum_{w \in q} \min\{ \mbox{\emph{freq}}(w, q), \mbox{\emph{freq}}(w, d) \}$,
        where, $\mbox{\emph{freq}}(w, q)$ and $\mbox{\emph{freq}}(w, d)$ are the counts of words in the question $q$ and the article $d$, respectively.
        Also, we add inverted document frequency to automatically scales down the importance of popular words.
        Furthermore, to make sure the computed scores are comparable across different questions, we normalize them to $[0, 1]$ by dividing $|q \cap q|$.

        \smallsection{Entity Features}
        We apply the spaCy\footnote{\url{http://spacy.io/}} toolkit to extract named entities from questions and articles.
        As both question and news article may contain multiple entities, we model them using the bag-of-entities representation.
        Analogous to the keyword features above, we can then compute their overlaps.

        \smallsection{word2vec Features}
        We utilize pre-trained word2vec 300-dimension vectors\footnote{GoogleNews-vectors-negative300.bin.gz from \url{https://code.google.com/archive/p/word2vec/}} and use the average vector to build vector representations for each question and news article.
        
        \smallsection{SVD Features}
        As an approximation, we use PCA analysis~\cite{dunteman1989principal} to determine the topics.
        More specifically, we first get the TF-IDF weighted bag-of-words representations of all articles after which we apply SVD decomposition to get the principal components.
        Finally, we project all questions and articles onto these components to get dense feature vectors.
        We further compute similarity based on these dense feature vectors, which indicates whether the news articles is related to the headline or not.

        Although we use similar features as the FNC winner (i.e., entity features are added and sentiment features are removed), we have achieved a substantially better classification results. 
        More than 30\% error reductions are observed in the relatedness classification in Section~\ref{subsec:relatedness-accuracy}, which demonstrates the importance of our newly designed \emph{entity features} based on Observation~\ref{conj:relatedness}.

    \subsection{Agreement Detection}

        In this section, we present our recurrent neural network (RNN) with attentions model designed for agreement categorization and document ranking within certain category.
        Although keyword/entity-based features work well for relevance classification, they cannot capture more subtle expressions that indicate agreement or disagreement.
        Recent advances on neural networks provide an automatic, high-quality way for this type of feature extraction.
        We design a RNN with attentions model for this purpose.
        
        While there are many variations of long-short term memory (LSTM), we use the following one for the rumor detection problem.
        Suppose the input sequence is $\X = (\x_1, \x_2, \ldots, \x_N)$, where $\x_k \in \mathbb{R}^l$ is the vector representation of the $k$-th element. At each position $k$, there is a set of internal vectors, including an input gate $\ig_k$, a forget gate $\fg_k$, an output gate $\og_k$, and a memory cell $\mc_k$. All these vectors together are used to generate a hidden state $\h_k \in \mathbb{R}^d$ as
        \begin{eqnarray*}
            \ig_k & = & \sigma(\W^i\x_k+\V^i\h_{k-1} + \mb^i) \\
            \fg_k & = & \sigma(\W^f\x_k+\V^f\h_{k-1} + \mb^f) \\
            \og_k & = & \sigma(\W^o\x_k+\V^o\h_{k-1} + \mb^o) \\
            \mc_k & = & \fg_k \odot \mc_{k-1} + \ig_k \odot \tanh(\W^c\x_k+\V^c\h_{k-1} + \mb^c) \\
            \h_k & = & \og_k \odot \tanh(\mc_k)
        \end{eqnarray*}
        where $\sigma$ is the sigmoid function, $\odot$ is the element-wise multiplication of two vectors, and all $\W^* \in \mathbb{R}^{d \times l}$, $V^* \in \mathbb{R}^{d \times d}$, and $b^* \in \mathbb{R}^d$ are parameters to be learned.

        Directly applying RNNs to model long articles is challenging.
        In order to capture and memorize useful information, RNNs require a bigger state size for the longer texts, and thus decrease efficiency.
        Fortunately, based on Observation~\ref{conj:stance}, it is possible to reduce long news articles to a few key sentences with only minimal loss of output quality.
        To obtain these sentences, we leverage word embeddings.
        Considering the limited training data and the model simplicity, we define the sentence embedding as the average of its pre-trained word embeddings.
        Specifically, we utilize the pre-trained Glove 300-dimension vectors and skip the stopwords when computing the average vector.
        Since questions usually consist of one or two sentences, we apply the same approach for them.
        We then evaluate the cosine similarity between the given question and all sentences in a news article.
        The sentences with the highest similarities to the question are the key sentences which then replace the news article text.
        The sentences are organized in their relative similarity order.
        In the following, we assume a default number of key sentences $k$ of 3.
        The effect of different values for $k$ will be discussed in Section~\ref{sec:param}.

        We follow Wang et al.~\cite{wang2015learning} to build a neural attention model, as shown in Figure~\ref{fig:rnn}.
        Formally, we have two sequences $\X^q = \{\x_1^q, \x_2^q, \ldots, \x_m^q\}$ and $\X^d = \{\x_1^d, \x_2^d, \ldots, \x_n^d\}$, where $m$ is the length of the question and $n$ is the number of tokens in the selected sentences, and each $\x$ is an embedding vector of the corresponding word.
        We build three LSTMs in total: \method{qLSTM} processes $\X^q$ and generates its hidden states $\h^q_{j}$; \method{dLSTM} reads $\X^d$ and outputs hidden states $\h^d_k$; and \method{mLSTM} models the matching between the question and the article and produces hidden states $\h^m_k$ which we discuss in detail later.

        Next, we generate the attention vectors $\att_k (1 \le k \le n)$ as follows.
        \begin{equation}
            \att_k = \sum_{j = 1}^{m} \alpha_{kj} \h_j^q
        \end{equation}
        Here, $\alpha_{kj}$ is an attention weight that encodes the degree to which $\x_k^d$ in the article is aligned with $\x_j^q$ in the question.

        The attention weight $\alpha_{kj}$ is generated as
        \begin{eqnarray}
            \alpha_{kj} & = & \frac{exp(e_{kj})}{\sum_{j'} exp(e_{kj'})} \\
            e_{kj}  & = & \w^e \cdot tanh(\W^q \h^q_j + \W^d \h^d_k + \W^m\h^m_{k-1})
        \end{eqnarray}
        where $\cdot$ is the dot product between two vectors and the vector $\w^e \in \mathbb{R}^d$ as well as all matrices $\W^{*} \in \mathbb{R}^{d \times d}$ are the parameters to be learned.

        The input of \method{mLSTM}, $\m_k$, is the concatenation of $\h^d_k$, which is the hidden state for the $k$-th token in the article, and $\att_k$, which is its attention weighted version.
        Thus, \method{mLSTM} will `remember' important matching results, and `forget' non-essential ones.

        To predict the agreement class of a news article, we use $\h^m_N$, i.e.,~the last hidden state of \method{mLSTM}.
        Instead of using a soft-max layer for 3-class classification, we choose to use two separate sigmoid modules for \emph{agree} and \emph{disagree}, which make the predicted scores comparable across different articles.
        % If neither of them have confidence scores above $0.5$, we output the stance to be \emph{discuss}.
        % Otherwise, we will predict the stance of a higher score and this score will later be used for ranking.

        Furthermore, we use an agreement score $\beta(q,d) \in [-1,+1]$ with $-1$ indicating maximum disagreement and $+1$ indicating maximum agreement.
        When score$_{\mbox{agree}}$ is larger than score$_{\mbox{disagree}}$, we let $\beta(q, d)$ be a positive score of score$_{\mbox{agree}}$. 
        Otherwise, we set $\beta(q, d)$ as a negative score of $-$score$_{\mbox{disagree}}$.
        Based on $\beta(q, d)$, we can define $P(y | q, d)$ accordingly as described in Section~\ref{sec:pre}.

    \subsection{Online Pipeline}
        Once an investigative question $q$ and its candidate collection $\mathcal{D}(q)$ arrive for processing, \our will first apply the tree-based model to compute the relatedness score $rel(q, d)$ for each article $d \in \mathcal{D}$.
        Then, for the articles with $rel(q, d) \ge 0.5$, \our will leverage the attention-based RNN to determine the agreement classes for each relevant news article.
        We will thus compute the agreement $\hat{y}$ based on $P(y | q, d)$.
        Note that at this stage, $P(y = \mbox{discuss} | q, d) = rel(q, d) \ge 0.5$.
        Therefore, if we finally get $\hat{y}$ as \emph{agree} or \emph{disagree}, its probability will be more than $0.5$.
        The \emph{agree} and \emph{disagree} articles will be ranked based on the absolute values of $\beta(q, d)$, while \emph{discuss} articles will be ranked by their $rel(q, d)$ scores.

%!TEX root = 00-main.tex
% !TEX encoding = UTF-8

\section{Experiments}
\label{sec:exp}

    Here we report the evaluation of \our on the real-world dataset.
   
    \subsection{Dataset}

        We evaluate \our on a recently published dataset, FNC-1\footnote{\url{https://github.com/FakeNewsChallenge/fnc-1}}, from the Fake News Challenge. 
        FNC-1 was designed as a stance detection dataset 
        %is used in an agreement detection challenge towards the fake new detection that attracts more than 50 teams all over the world in the summer of 2017.
        and it contains 75,385 labeled headline and article pairs. 
        The labels are analogous to the agreement classes that we consider, namely \emph{agree}, \emph{disagree}, \emph{discuss}, and \emph{unrelated}.
        Each headline in the dataset is phrased as a statement.
        Note that our techniques hold for statements as well as investigative questions.
        In fact, we observe that investigative questions are most commonly rephrased statements.
        %In fact, \our also works for the statement, while the question are a special type of the statement.
        %Also, transforming these headlines into investigative questions will not make much difference.
        %In many cases you can do a reasonable transformation by taking headline X and writing it as "Is it true that X?", e.g., ``Is it true that Robert Plant ripped up \$800M Led Zeppelin reunion contract?''
        %So we treat these headlines as questions directly.
        Detailed statistics of the dataset can be found in Table~\ref{tbl:dataset}.

        Note furthermore that the topics mentioned in the questions and articles in the training and testing sets are significantly different.
        Consequently, this setting is challenging and even harder than a real-world setup where partial overlap can often be assumed.

    \subsection{Evaluation Metrics}
    
        Since some of the questions in this dataset are not controversial, we present evaluation results in two folds: (1) all questions and (2) controversial questions.
        For both, we evaluate all compared methods using the following three metrics: (1) NDCG@$K$ and Avg.\ NDCG for the ranking accuracy, (2) relatedness accuracy for the classifier's performance, and (3) the official FNC metric, weighted accuracy.
        Considering the \our's interface as shown in Figure~\ref{fig:interface}, we think the NDCG@$K$ and Avg.\ NDCG is the most important.
        Details are as follows.
       
        \smallsection{NDCG@$K$ and Avg.\ NDCG}
        Because we are presenting three ranked lists of articles to the user, we utilize the normalized discounted cumulative gain, NDCG@$K$, for each investigative question and calculate the average over all questions for evaluation.
        
        The gain of an article in a ranked list is defined as follows.
        In the ranked list of label \emph{agree}, only \emph{agree} articles will receive a score of 1, while other articles will get a zero score.
        Articles in the \emph{disagree} and \emph{discuss} list are treated analogously.
        % In the \emph{discuss} list, all \emph{related} (\emph{agree}, \emph{disagree}, and \emph{discuss}) articles will get a score of 1, while \emph{unrelated} articles will receive a zero score.

        Given a question and a ranked list of $K$ articles, the discounted cumulative gain is calculated as
        $$
            \mbox{DCG@}K = gain_1 + \sum_{i = 2}^{K} \frac{gain_i}{\log_2(i)}
        $$
        The NDCG@$K$ is then computed as a normalization by the best possible DCG@$K$. 
        If the ideal DCG@$K$ is 0 for any of the lists, we will skip this ranked list for this question.
        Considering the numbers of articles from each class displayed in our proposed interface (\ie, Figure~\ref{fig:interface}), we evaluate NDCG@$3$ for both \emph{agree} and \emph{disagree} ranked lists, and NDCG@$5$ for the \emph{discuss} ranked list. 

        Since all questions as well as their three ranked lists are equally important for presenting the holistic view towards the investigated question to the user, to conduct an overall comparison, we define the average NDCG score as follows. 
        For each question, we first average NDCG scores of all three ranked list. 
        \textbf{Avg. NDCG} is computed as the average of these averages for different questions.

        % \smallsection{Relatedness Accuracy}
        % In the relatedness accuracy, we consider only two classes: \emph{related} \emph{vs}.\ \emph{unrelated}.
        % The score is then calculated by articles (not) matching their underlying class correctly.
        \smallsection{Relatedness Error}
        To evaluate the relatedness classifier, we consider only two classes: \emph{related} \emph{vs}.\ \emph{unrelated}.
        The relatedness error refers to the percentage of misclassified question-article pairs.

        \smallsection{Weighted Accuracy}
        This is the official metric for FNC-1: For a question and an article, if the model successfully predicts the \emph{related}/\emph{unrelated} label, it receives a score of $0.25$. 
        For a question and a \emph{related} article, if the model successfully predicts \emph{agree}, \emph{disagree}, or \emph{discuss}, it receives a score of $0.75$. 
        The final score is then normalized by the maximum possible score\footnote{For more details, please refer to \url{http://www.fakenewschallenge.org/}}.

    \subsection{Experimental Setting}
        All experiments are conducted on a single machine equipped with an Intel Xeon processor E5-2650@2.2GHz and a NVIDIA GeForce GTX 1080. 
        In \our, the tree-based model is implemented in XGBoost~\cite{chen2016xgboost} and the RNN+attention model is implemented using Tensorflow~\cite{abadi2016tensorflow}. 
        The source code is available in the authors' GitHub\footnote{\url{https://github.com/shangjingbo1226/Maester}}.

        \smallsection{Maester}
        This is our proposed model.
        By default, the number of key sentences, $k$, is set to $3$, and the number of training epochs is set to $10$. 
        For further details on the parameters, please refer to the study on parameter sensitivities in Section~\ref{sec:param}.
        As our models contain some randomness, we run all experiments five times and report the average performance.

        \smallsection{FNC-1 Winner}
        As we discussed before, the FNC-1 winner's solution is an ensemble of a tree-based and a convolutional neural network (CNN) models.
        This combined model is able to detect the relatedness of the article effectively, primarily due to their effective tree-based model with human designed features like TF-IDF weighted keywords.
        However, it is limited in detecting the actual \emph{agree} or \emph{disagree} label of articles.
        Since the dataset is imbalanced, most of the related articles are labelled \emph{discuss} and \emph{disagree} labels are rare.
        Thus, the winner's solution will aggressively classify most of articles as \emph{discuss} and the rest as \emph{agree}, in order to achieve a high overall accuracy.
        However, this leads to a poor ranking performance.
        % In the following, we use \textbf{FNC Winner (Tree)} and \textbf{FNC Winner (CNN)} to denote the tree-based model and the CNN model in FNC-1 winner's original solution respectively.
        We report the best performance for \method{FNC-1 Winner} during the competition.

        \smallsection{Alternative Models}
        As an alternative to our two-step framework, we also considered more straightforward models that have been applied in similar use cases before.
        The first of these is \textbf{bag-of-words}.
        It is unsuitable for our use case as language is evolving and there may be different vocabulary present in the application than in the training data.
        However, combining bag-of-words with some feature selection techniques leads to some interesting keywords that signal different types of agreement.
        For example, we observe that ``reportedly'' is a strong signal for \emph{discuss}.
        We tried incorporating keyword lists based on the bag-of-words model in our own framework, however, improvements were negligible.
        Another type of models that is widely adopted when learning to match questions and articles is \textbf{matrix factorization}~\cite{shang2014parallel}.
        In our experiments, we observed that this technique has worse and unstable performance for this particular problem.
        Again, this is caused by the fact that not all words appearing in the application or test dataset are covered in the training data.
        %The performance of these methods are still far below our expectation. 
        For example, the weighted accuracy of the bag-of-words model is only $77.64\%$. 
        The weighted accuracy of the matrix factorization approach is similar.
        Therefore, they are not included in this evaluation.
        
    \subsection{Relatedness Error}\label{subsec:relatedness-accuracy}

        % \begin{table}[t]
        %     \caption{Error rate of relatedness classification. More than $30\%$ of error reductions are achieved by \our over FNC-1 Winner.}\label{tbl:coarse-grained-stance-detection}
        %     \vspace{-0.3cm}
        %     \begin{tabular}{|c|c|c|}
        %         \hline
        %               % & \multicolumn{2}{|c|}{Accuracy} \\
        %         % \hline
        %         Method & All Questions  & Controversial Questions \\
        %         \hline
        %         \hline
        %         FNC-1 Winner &   3.04\%    & 3.75\% \\
        %         % FNC-1 Winner &   96.96\%    & 96.25\% \\
        %         % \hline
        %         % FNC-1 Winner (Tree) &   97.70\%    & 97.30\% \\
        %         % \hline
        %         % FNC-1 Winner (CNN) &  76.96\%     & 70.01\%  \\
        %         \hline
        %         \our &   \textbf{2.13\%}    & \textbf{2.46\%} \\
        %         % \our &   \textbf{97.87\%}    & \textbf{97.54\%} \\
        %         \hline
        %     \end{tabular}
        % \end{table}
        
        \begin{table}[!t]
            \caption{Error rate of relatedness classification. More than $30\%$ of error reductions are achieved by \our over FNC-1 Winner.}\label{tbl:coarse-grained-stance-detection}
            \vspace{-0.3cm}
            \scalebox{0.95}{
                \begin{tabular}{ccc}
                    \toprule
                    Method & All Questions  & Controversial Questions \\
                    \midrule
                    FNC-1 Winner &   3.04\%    & 3.75\% \\
                    \our &   \textbf{2.13\%}    & \textbf{2.46\%} \\
                    \bottomrule
                \end{tabular}
            }
            \vspace{-0.3cm}
        \end{table}

        We first study \our's performance on the relatedness classification task. 
        As shown in Table~\ref{tbl:coarse-grained-stance-detection}.
        % Interestingly, \method{FNC-1 winner (Tree)} achieves better performance than \method{FNC-1 Winner}, independent of the question specifics.
        % Therefore, we argue that the effectiveness of \method{FNC-1 Winner} when it comes to relatedness classification is mainly due to its tree-based model.
        \our has the best performance and achieves more than $29.93\%$ and $34.40\%$ error reductions on all questions and controversial questions, respectively.
        This demonstrates the importance of the added entity features compared to previously utilized sentiment features which tend to be noisy.
        An error rate less than $3\%$ demonstrates that \our's tree-based model built upon handcrafted features is precise enough to predict whether a document is related or not.

        To compare the significance of different features, we calculate the relative feature importance for each feature type using the built-in function in XGBoost~\cite{chen2016xgboost}, as shown in Table~\ref{tbl:feature-importance}.
        Here, we can see that the combined importance of keyword features and entity features is significant, i.e.,~52.18\%.
        Moreover, the newly added entity features are more important than the word2vec and SVD features.
        % Also, word2vec and SVD features are kind of implicit keywords and entity features.
        Therefore, Observation~\ref{conj:relatedness} has been verified with this experiment.

\begin{table*}[!htb]
    \begin{minipage}{0.25\textwidth}
        \caption{Feature importance.}\label{tbl:feature-importance}
        \vspace{-0.3cm}
        \begin{tabular}{cc}
            \toprule
            \textbf{Feature} & \textbf{Importance} \\
            \midrule
            Keyword & 29.68\% \\
            Entity & 22.50\% \\
            word2vec &  13.75\% \\
            SVD & 34.07\% \\
            \bottomrule
        \end{tabular}
    \end{minipage}%
    \begin{minipage}{0.75\textwidth}
        \caption{Ranking performance of the agreement-aware search framework.} \label{tbl:ranking-performance}
        \vspace{-0.3cm}
        \small
\scalebox{0.915}{
        \begin{tabular}{ccccccccc}
            \toprule
            \multirow{3}{*}{\textbf{Method}} & \multicolumn{4}{c}{\textbf{All Questions}} & \multicolumn{4}{c}{\textbf{Controversial Questions}}\\
            \cmidrule{2-9}
             & Agree & Disagree & Discuss & \textbf{Avg.} & Agree & Disagree & Discuss & \textbf{Avg.} \\
             & NDCG@3 & NDCG@3 & NDCG@5 & \textbf{NDCG} & NDCG@3 & NDCG@3 & NDCG@5 & \textbf{NDCG}\\
            \midrule
            FNC-1 Winner & \textbf{51.71}\% & 2.31\% & 64.04\% & 39.38\% & \textbf{43.75}\%   & 2.58\%  & 31.90\% & 26.08\% \\
            \midrule
            \our &  48.11\%     & \textbf{20.38\%} & \textbf{68.20\%} & \textbf{47.62\%} &  40.88\%     & \textbf{19.13\%} & \textbf{61.39\%} & \textbf{40.47\%} \\
            \bottomrule
        \end{tabular}
}
    \end{minipage}
    \vspace{-0.4cm}
\end{table*}
        
    \subsection{Ranking Evaluation}

        We evaluate the results as three ranked lists.
        This ranking evaluation is crucial because our ultimate goal is to present a holistic view towards the user's question.
        
        As shown in Table~\ref{tbl:ranking-performance}, \our achieves the best overall agreement-aware ranking performance. 
        \our's Avg. NDCG score is much higher than \method{FNC-1 Winner}'s Avg. NDCG score, for both controversial and non-controversial questions. 
        Specifically, for controversial questions, \our's almost doubles \method{FNC-1 Winner}'s performance, while for both controversial and non-controversial questions, the improvement is $20\%$.
        We also notice that \emph{disagreement} class is the most challenging one among all the three classes, and \our achieves a 7-fold improvement for this class.

        % The experimental results of ranking performance are summarized in Table~\ref{tbl:ranking-performance}.
        % As the overall agreement-aware ranking measurement, \our's Avg NDCG is much higher compared to \method{FNC-1 Winner}'s, independent whether all types or only controversial questions are evaluated.
        % Specifically, for controversial questions, \our's improvement is about 2x; over both controversial and non-controversial questions, the improvement is $20\%$.
        % Moreover, for controversial questions, \our achieves a 7-fold improvement in the especially difficult case of disagreement.
        The improvements on the NDCG score in the \emph{discuss} class are also noticeable. 
        The NDCG score in the \emph{agree} class is slightly lower than the reference score but is still comparable.
        These significant ranking improvements demonstrate that \our is a better fit than \method{FNC-Winner} as a helpful rumor news investigation tool.
        
        Finally, from this ranking evaluation, we obtain a better understanding about the \method{FNC-1 Winner}. 
        It achieves the high weighted accuracy through aggressively predicting articles as \emph{agree} and \emph{discuss} where very few articles are categorized as \emph{disagree}.
        However, such biased prediction gets punished when evaluating ranking performance.

    \subsection{FNC metric: Weighted Accuracy}

        % The task of agreement classification is more challenging than relatedness computation as it requires the model to distinguish articles in a more subtle way.
        Since \method{FNC-1 Winner} is specifically optimized for the official metric (i.e., weighted accuracy) in the challenge, we also used the weighed accuracy for evaluation.
        % We adopt weighted accuracy to put more weights on \emph{agree}, \emph{disagree}, and \emph{discuss}.
        From Table~\ref{tbl:fine-grained-stance-detection},
        we can find that \our outperforms \method{FNC-1 winner} where the absolute improvement of accuracy is $0.96\%$ and $2.88\%$ on all questions and controversial questions, respectively.
        Considering that \method{FNC-1 winner} has won the FNC by a margin of $0.05\%$, these improvements can be considered as remarkable.

        In fact, recall that \our relies only on the top-$3$ key sentences from the article, whereas \method{FNC-1 Winner} considers all sentences in the article.
        These results reflect that using only three key sentences can still capture enough information to detect agreement.
        % Therefore, Observation~\ref{conj:stance} is reasonable.

        % \begin{table}[t]
        %     \caption{Weighted accuracy of agreement detection. Note that \method{FNC-1 winner} wins the challenge by an advantage of $0.05\%$. \our's improvements should be considered as significant.} \label{tbl:fine-grained-stance-detection}
        %     \vspace{-0.3cm}
        %     \begin{tabular}{|c|c|c|}
        %         % \hline
        %               % & \multicolumn{2}{|c|}{Weighted Accuracy} \\
        %         \hline
        %         Method & All Questions  & Controversial Questions  \\
        %         \hline
        %         \hline
        %         FNC-1 Winner &    82.02\%          & 66.66\% \\
        %         \hline
        %         % FNC-1 Winner (Tree) &       82.98\%  &  67.20\%  \\
        %         % \hline
        %         % FNC-1 Winner (CNN) &       60.88\%   & 49.32\%   \\
        %         % \hline
        %         \our  &   \textbf{82.98\%}  & \textbf{69.54\%} \\
        %         \hline
        %     \end{tabular}
        % \end{table}
        
        \begin{table}[t]
            \caption{Weighted accuracy of agreement detection. Note that \method{FNC-1 winner} wins the challenge by an advantage of $0.05\%$. \our's improvements should be considered as remarkable.} \label{tbl:fine-grained-stance-detection}
            \vspace{-0.3cm}
            \begin{tabular}{ccc}
                \toprule
                % \hline
                       % & \multicolumn{2}{|c|}{Weighted Accuracy} \\
                Method & All Questions  & Controversial Questions  \\
                \midrule
                FNC-1 Winner &    82.02\%          & 66.66\% \\
                \hline
                % FNC-1 Winner (Tree) &       82.98\%  &  67.20\%  \\
                % \hline
                % FNC-1 Winner (CNN) &       60.88\%   & 49.32\%   \\
                % \hline
                \our  &   \textbf{82.98\%}  & \textbf{69.54\%} \\
                \bottomrule
            \end{tabular}
            \vspace{-0.3cm}
        \end{table}

\begin{table*}[t]
    \caption{Top-3 key sentences determined by \our for agreement detection.}\label{tbl:case_study}
    \vspace{-0.2cm}
    \centering
    \small
    \begin{tabularx}{1.0\textwidth}{lp{15cm}}
        \toprule
        Question & Is it true that a woman pays \$20,000 for third breast to make herself LESS attractive to men? \\
        \midrule
        \midrule
        \multirow{3}{*}{An \emph{agree} article} & 1. No, you do not need to adjust your sets, you are actually looking at a woman with three breasts.  \\
        & 2. Jasmine added: I got it \textbf{because I wanted to make myself unattractive to men}. \\
        & 3. She denies that she had the extra breast put on to get fame and fortune. \\
        \midrule
        \multirow{3}{*}{A \emph{disagree} article} & 1. Did a woman claiming to have a third breast play a hoax on us? \\
        & 2. A top plastic surgeon, Mr Nilesh Sojitra, also cast doubt over the surgery after claiming \textbf{no reasonable doctor would perform the operation}. \\
        & 3. Snopes.com came up with a number of intriguing arguments that could indicate Jasmine Tridevil \textbf{did not actually pay \$20,000 for an extra breast.} \\
        \bottomrule
    \end{tabularx}
    \vspace{-0.3cm}
\end{table*}

    \subsection{Parameter Sensitivities} \label{sec:param}

        Here, we study the parameter sensitivities for the two major parameters in \our: (1) the number of key sentences, $k$ and (2) the number of epochs needed for model convergence.

\begin{figure}[t]
            \centering
            \subfigure[On all questions.]{
                \includegraphics[width = 0.22\textwidth]{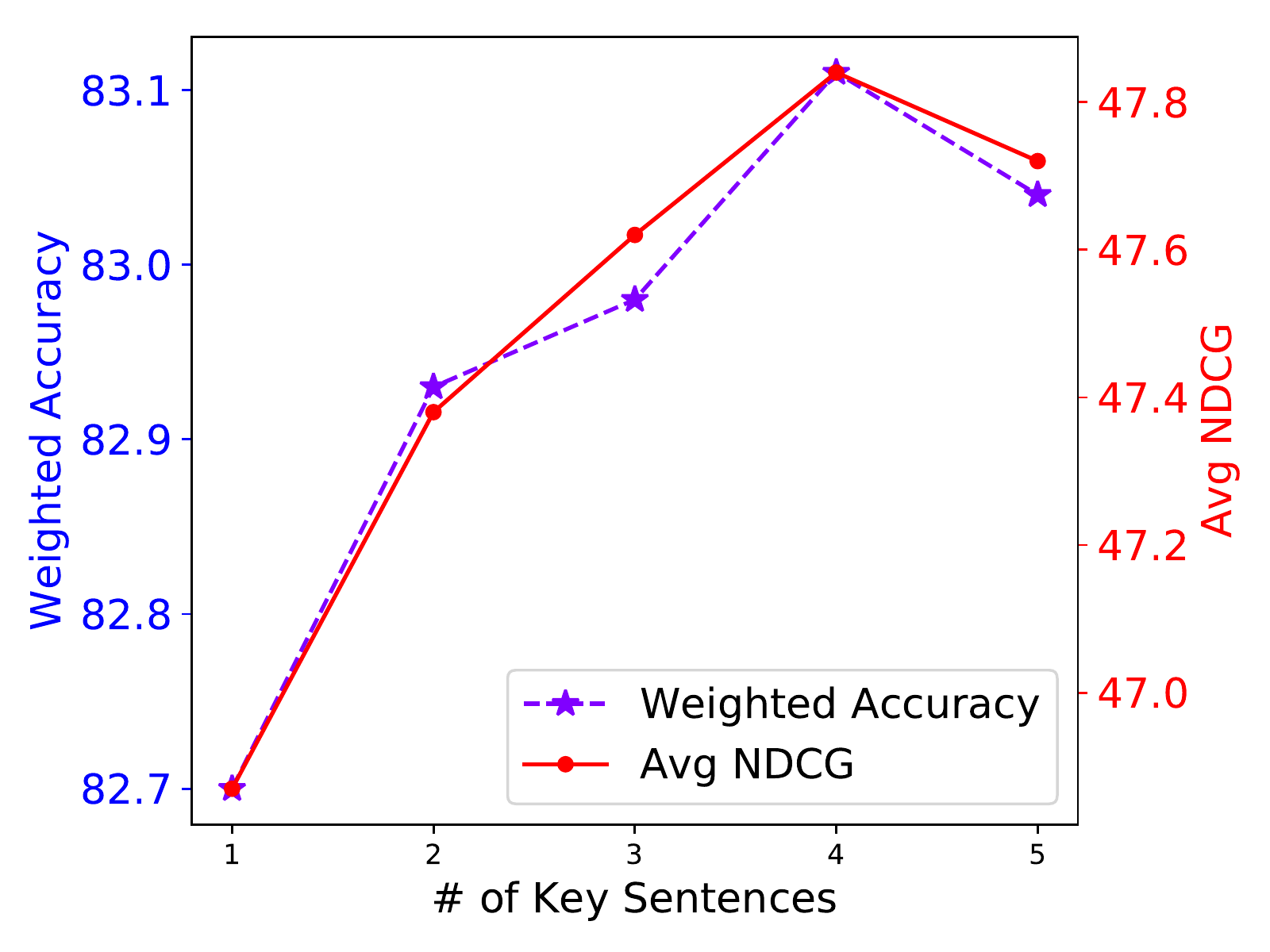}
            }
            \subfigure[On controversial questions.]{
                \includegraphics[width = 0.22\textwidth]{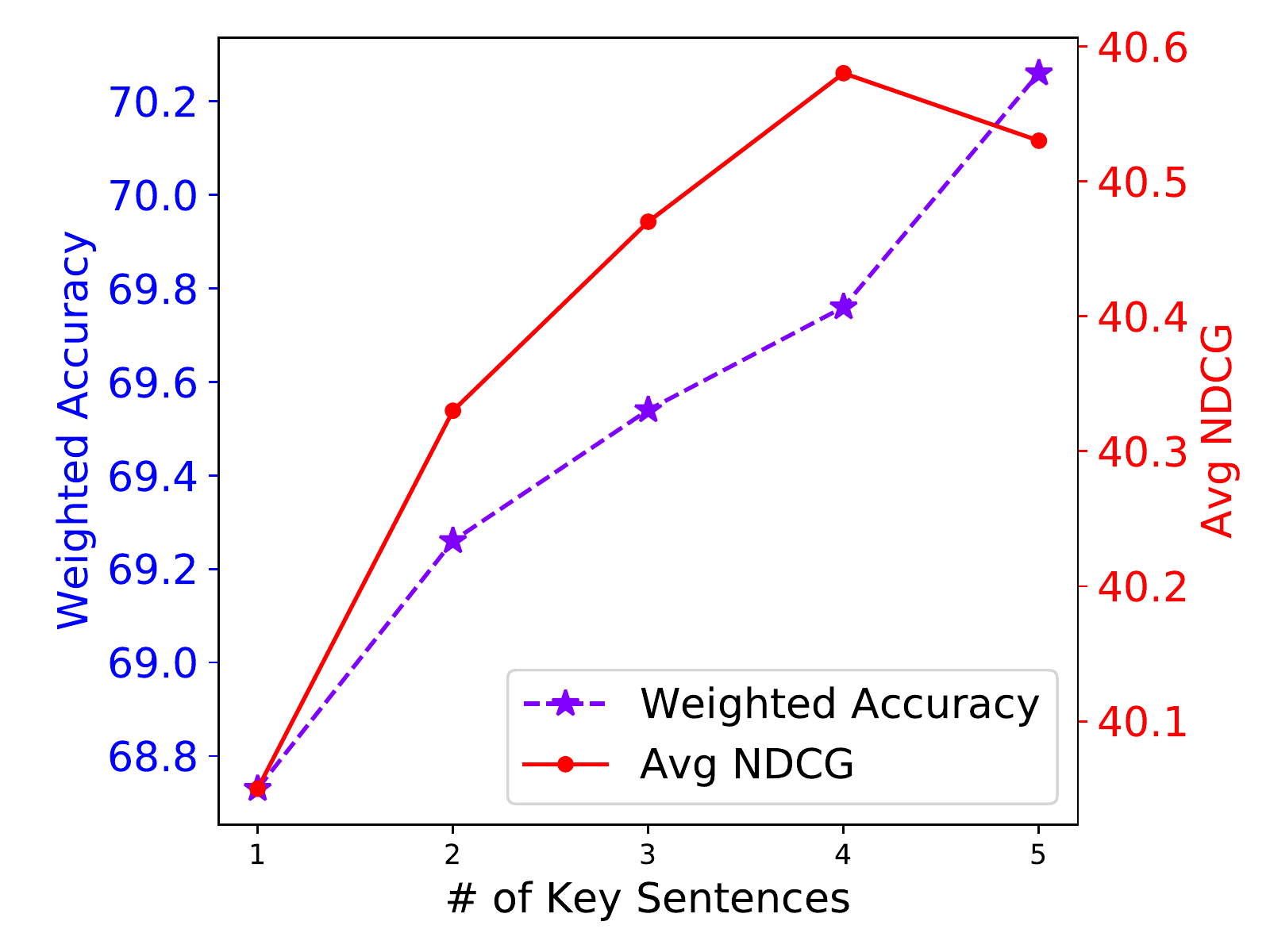}
            }
            \vspace{-0.4cm}
            \caption{How many key sentences are enough?}
            \label{fig:key_sentences}
            \vspace{-0.5cm}
        \end{figure}

        As shown in Figure~\ref{fig:key_sentences} only knowing the top sentence of an article already provides good quality results.
        When more key sentences are available, the weighted accuracy on controversial questions grows constantly, while the ranking performance drops a little when $k=5$ is reached.
        This implies that more sentences disclose more information, however, a few key sentences are enough for good ranking quality, which supports Observation~\ref{conj:stance}.
        %Therefore, the default setting of $k=3$ is reasonable, although $k=4$ could lead to a slightly better performance.

\begin{figure}[t]
            \subfigure[On all questions.]{
                \includegraphics[width = 0.22\textwidth]{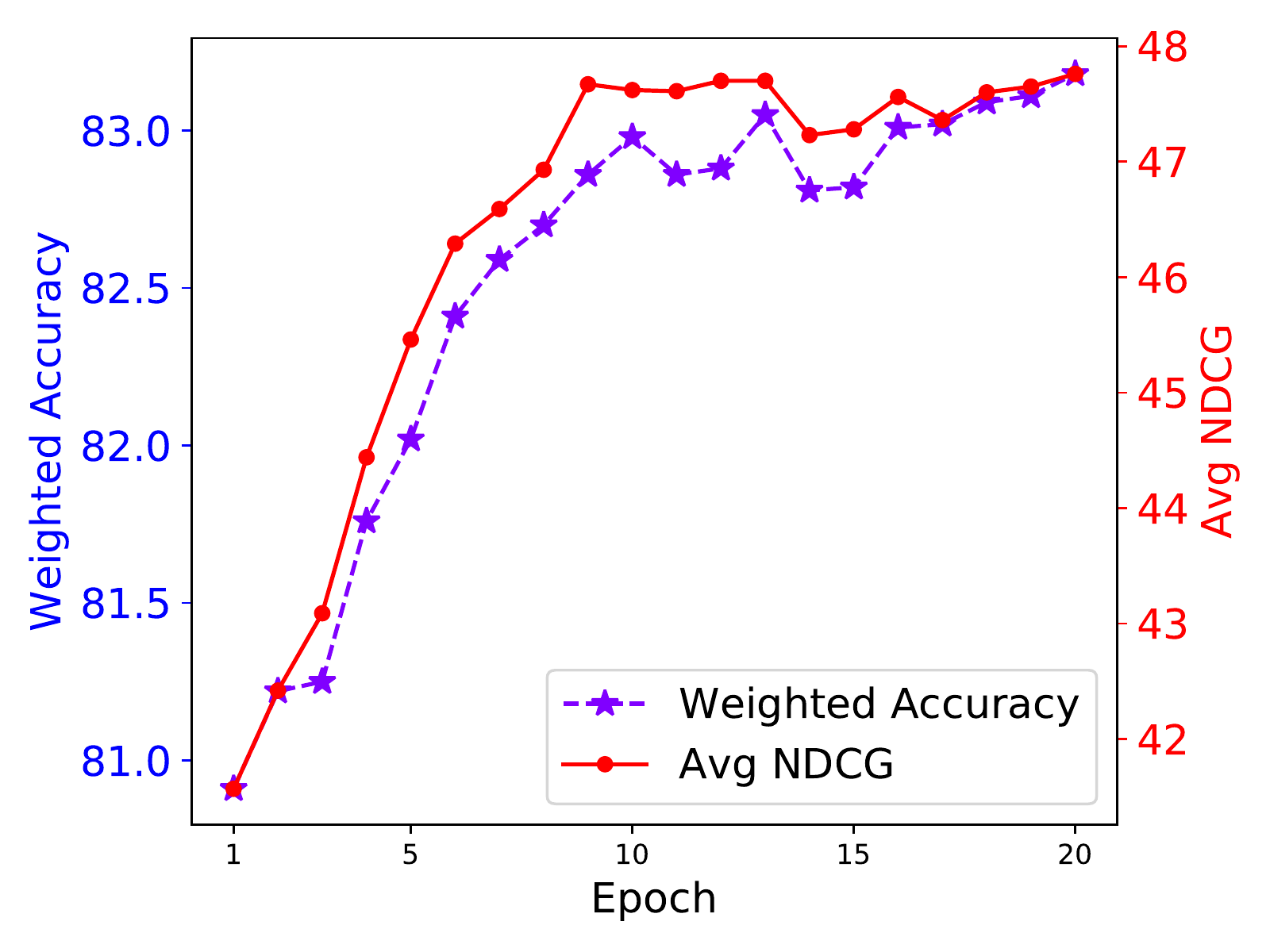}
            }
            \subfigure[On controversial questions.]{
                \includegraphics[width = 0.22\textwidth]{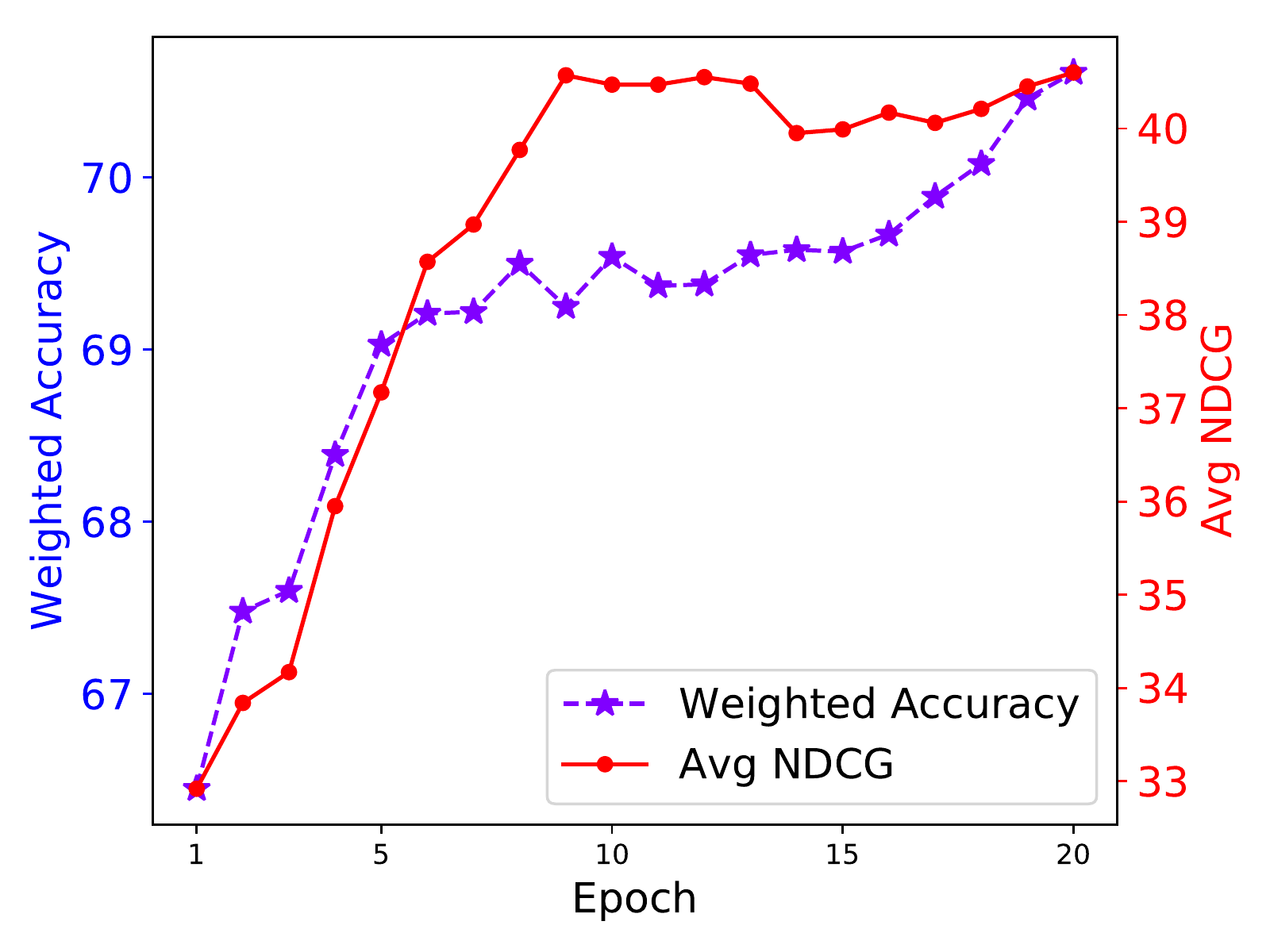}
            }
            \vspace{-0.4cm}
            \caption{Convergence study on test data.}
            \label{fig:epoch}
            \vspace{-0.5cm}
        \end{figure}

        Second, we studied the convergence of the RNN+attention model in \our in Figure~\ref{fig:epoch}. 
        The results show that the result quality, measured with either weighted accuracy or Avg NDCG, stabilizes after 10 epochs.
        This is a promising time span for early stops and savings on training time.

    \vspace{-0.1cm}
    \subsection{Efficiency Evaluation}

        The whole \our pipeline, including both tree-based and RNN models, can be trained within 1 hour.
        However, in a real-world application, online serving time is more important.
        \our can process a pair of question and article within about 5.86 ms.
        Specifically, in our setup, \our spends about 0.16 seconds on average to present the final results (as shown in Figure~\ref{fig:interface}) to the user.

    \vspace{-0.1cm}
    \subsection{Case Study}
        For a controversial question, we randomly pick two articles from the \emph{agree} and \emph{disagree} classes and show the top-$3$ key sentences selected by \our in Table~\ref{tbl:case_study}.
        From these results, we observe that the chosen sentences, especially the highlighted parts, are essential for agreement classification.
        Moreover, for this question, \our achieves $100\%$ NDCG@3 in both \emph{agree} and \emph{disagree} ranked lists, while the \method{FNC-1 winner}'s scores are $29.82\%$ and $0\%$, respectively.
        These findings further consolidate our Observation~\ref{conj:stance}.
        
% \begin{table}[t]
%     \caption{Case Study: Key sentences as determined by \our for agreement detection.}\label{tbl:case_study}
% \scalebox{0.79}{
%     \begin{tabular}{|p{1.3cm}|p{8.5cm}|}
%         \hline
%         Question & Is it true that a woman pays \$20,000 for third breast to make herself LESS attractive to men? \\
%         \hline
%         \hline
%          & Our Selected Top-3 Similar Sentences \\
%         \hline
%         An \emph{agree} article  & 1. No, you do not need to adjust your sets, you are actually looking at a woman with three breasts.  2. Jasmine added: I got it \textbf{because I wanted to make myself unattractive to men}. 3. She denies that she had the extra breast put on to get fame and fortune. \\
%         \hline
%         A \emph{disagree} article & 1. Did a woman claiming to have a third breast play a hoax on us? 2. A top plastic surgeon, Mr Nilesh Sojitra, also cast doubt over the surgery after claiming \textbf{no reasonable doctor would perform the operation}. 3. Snopes.com came up with a number of intriguing arguments that could indicate Jasmine Tridevil \textbf{did not actually pay \$20,000 for an extra breast.} \\
%         \hline
%     \end{tabular}
% }
% \end{table}
% !TEX encoding = UTF-8
%!TEX root = 00-main.tex

\section{Conclusion \& Future Work}\label{sec:con}

In this paper, we focus on investing rumor news using an agreement-aware article search. 
%Solving this problem is beneficial to both individual users and society as a whole.
We develop an agreement-aware search framework that is designed to provide users with a holistic view of an investigative question, for which the ground truth is not certain.
Based on two intuitive but important observations, we designed a two-step model consisting of a tree-based model based on handcrafted features and a RNN+attention model focusing on only a few key sentences.
Our experimental results and case studies not only demonstrate the effectiveness of our model, but also verify both observations empirically.

There are many related problems and follow-up work that should be explored in the future.
In the context of rumor detection, we propose using statements, here in the form of controversial questions, to further the understanding of a topic.
However, it remains unclear how to derive such statements.
Another line of interesting follow-up work is to allow not only a limited set of labels but to enable additional entity-driven options.
For example, given the question ``Who is the best basketball player in history?'' many people will say ``Michael Jordan'' but there are others who will mention names such as ``Kobe Bryant'' and ``Lebron James''.

\section*{Acknowledgements}

Research was sponsored in part by U.S. Army Research Lab. under Cooperative Agreement No. W911NF-09-2-0053 (NSCTA), DARPA under Agreement No. W911NF-17-C-0099, National Science Foundation IIS 16-18481, IIS 17-04532, and IIS-17-41317, DTRA HDTRA11810026, Google PhD Fellowship, and grant 1U54GM114838 awarded by NIGMS through funds provided by the trans-NIH Big Data to Knowledge (BD2K) initiative (www.bd2k.nih.gov). Any opinions, findings, and conclusions or recommendations expressed in this document are those of the author(s) and should not be interpreted as the views of any U.S. Government. The U.S. Government is authorized to reproduce and distribute reprints for Government purposes notwithstanding any copyright notation hereon.

\bibliographystyle{acm}
\vspace{-0.1cm}
{\normalsize
\bibliography{cited}}

\end{document}